\documentclass[epj,final]{svjour}
\usepackage{amssymb}
\usepackage{epsfig}
\newcommand{\non}{\nonumber}
\begin{document}
\title{Longitudinally Polarized Photoproduction of Inclusive Hadrons at Fixed-Target Experiments}
\author{B.\ J\"{a}ger\inst{1} \and M.\ Stratmann\inst{2} \and W.\ Vogelsang\inst{3}\inst{4}
%
}                     
\institute{Institut f{\"u}r Theoretische Physik, Universit{\"a}t Karlsruhe,
D-76128 Karlsruhe, Germany \and
Institut f{\"u}r Theoretische Physik, Universit{\"a}t Regensburg,
D-93040 Regensburg, Germany \and
Physics Department, Brookhaven National Laboratory,
Upton, New York 11973, U.S.A. \and
RIKEN-BNL Research Center, Brookhaven 
National Laboratory, Upton, New York 11973 -- 5000, U.S.A.}
%
\date{}
%
\abstract{
We present a detailed phenomenological study of spin-dependent single-inclusive 
high-$p_T$ 
hadron photoproduction with particular emphasis on the kinematics relevant for 
the {\sc Compass} and {\sc Hermes} fixed-target experiments. 
We carefully examine the theoretical uncertainties
associated with the only moderate 
transverse momenta accessible in such measurements and
analyze the sensitivity of the relevant spin asymmetries to 
the gluon polarization in the nucleon as well as to the 
completely unknown parton content of circularly polarized photons.
} 
\maketitle

\vspace*{-13cm}
\noindent
BNL-NT-05/12, RBRC-510, SFB/CPP-05-15
\vspace*{11.2cm} 
%
\section{Motivation and Introduction}
\label{intro}
%
To measure the so far largely unknown polarization of gluons, $\Delta g$, in
the nucleon is a key goal of several current experiments 
in high-energy nuclear physics. 
The successful start of polarized proton-proton collisions at
the BNL Relativistic Heavy Ion Collider (RHIC) has opened up new, unequaled 
possibilities~\cite{ref:rhic}. 
Gluon polarization can be accessed in a variety of high transverse 
momentum processes such as single-in\-clu\-sive hadron,
jet, prompt photon, or heavy quark production. 
{\sc Compass} \cite{ref:compass} at CERN and {\sc Hermes}  \cite{ref:hermes} 
at DESY, instead, scatter beams of longitudinally polarized leptons
$l$ off polarized fixed targets $N$. Here, in particular 
high-$p_T$ hadron pairs, both in photoproduction 
and in deep-inelastic electroproduction, have been identified by the 
experiments to provide 
a promising tool to gain some knowledge about $\Delta g$ \cite{ref:twohadron}.

A meaningful extraction of parton densities from experiment requires a reliable 
interpretation of the underlying data in the framework of perturbative QCD (pQCD).
Factorization theorems \cite{ref:fact} assure that in the presence of a large scale, for
instance, a high transverse momentum $p_T$, a cross section can be
written as a convolution of partonic hard scattering cross sections, which
depend on the process under consideration, certain combinations
of non-perturbative, {\em universal} parton densities, and, if applicable, 
fragmentation functions.
The standard pQCD framework receives corrections which are suppressed by
inverse powers of the large scale characterizing the process.
Since it is not a priori clear where these ``power corrections'' become
relevant, the factorized ansatz has to be scrutinized for each measurement.
For all spin-dependent observables, cross sections or spin asymmetries, 
this is best achieved by studying {\em first}
the unpolarized cross sections because both ingredients, the parton
densities and the hard scattering cross sections, are known very well here.
Only if a reasonable agreement between theory and experiment is
established within their respective uncertainties
can one have confidence that a similar measurement with polarization
can be used to extract information about the spin structure of the nucleon.

\begin{figure*}[ht]
\vspace*{-0.5cm}
\begin{center}
\begin{minipage}{7cm}
\epsfig{figure=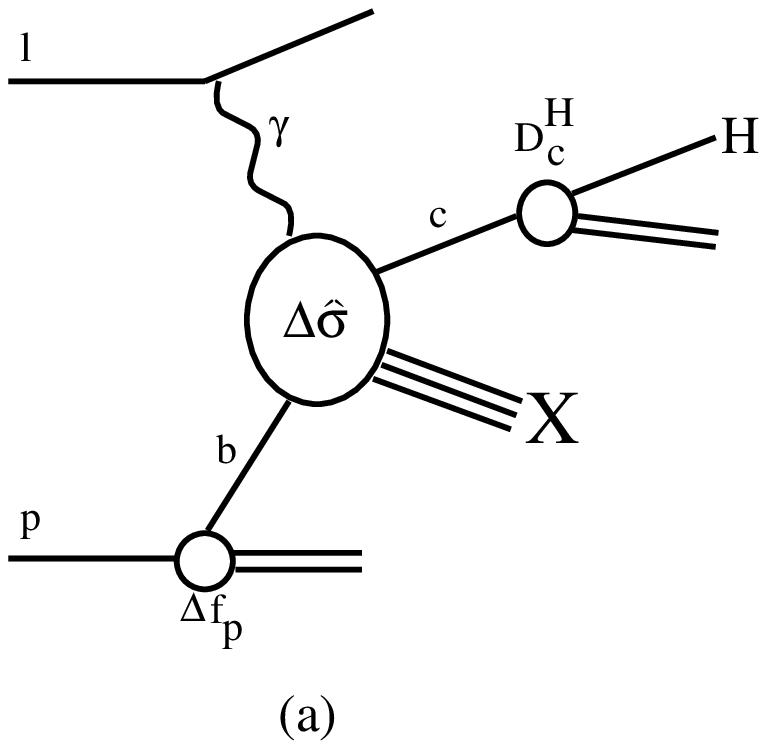,width=0.85\textwidth}
\end{minipage}
\begin{minipage}{7cm}
\epsfig{figure=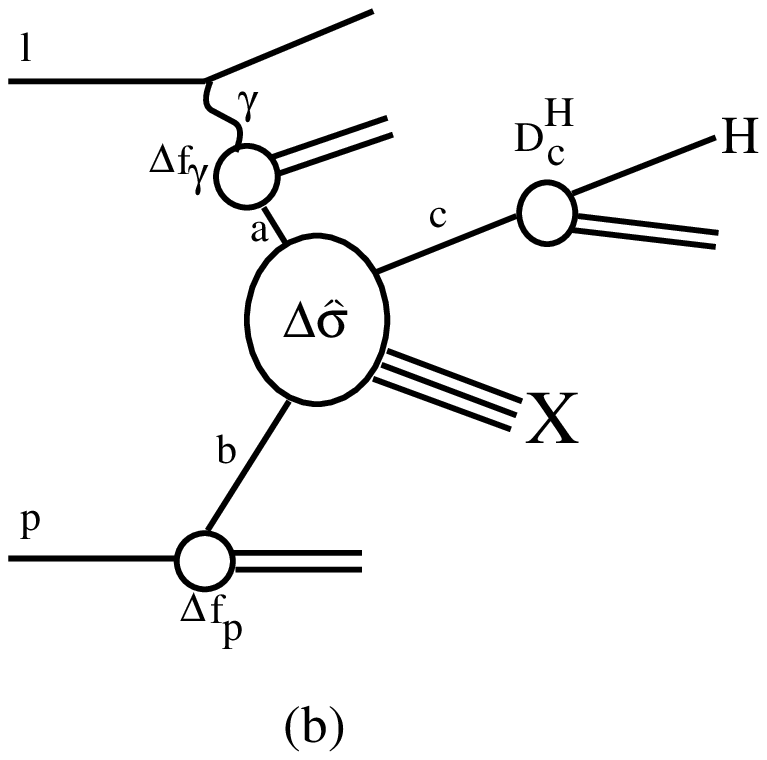,width=0.85\textwidth}
\end{minipage}
\end{center}
\vspace*{-0.5cm}
\caption{\sf Generic direct (a) and resolved (b) photon contributions to
the photoproduction of a hadron $H$. \label{fig:cartoon}}
\end{figure*}
First measurements at RHIC of {\em un}polarized high-$p_T$ neutral pion 
\cite{ref:rhic-unppion} and prompt photon \cite{ref:rhic-unpphoton} 
production rates turned out to be in good agreement with theoretical
predictions \cite{ref:pionnlo,ref:pionothers,ref:photonnlo}
even down to unexpectedly small values of the transverse
momentum of about $1.5\,\mathrm{GeV}$. For this comparison it is
essential to compute the perturbatively calculable short-distance 
cross sections up to the next-to-leading order (NLO) approximation of pQCD.
On the one hand, NLO corrections turn out to be sizable, on the other
hand, theoretical uncertainties related to the renormalization and
factorization procedures are much reduced compared to lowest order (LO)
estimates. Therefore much effort was put in calculating the
NLO pQCD corrections also for spin-dependent processes in recent years.

The {\sc Phenix} collaboration at RHIC has recently released first,
partly preliminary, results for high-$p_T$ neutral pion production with
longitudinally polarized protons \cite{ref:pionphenix}. However, from this 
measurement alone it is difficult and way too early to draw any conclusions 
about the gluon polarization $\Delta g$ in the nucleon; see also
\cite{ref:pionprl}. To map out $\Delta g(x,\mu)$
over a sufficiently large range in the momentum fraction $x$, which is
ultimately required for a determination of the first moment of
$\Delta g$, and scale $\mu$ (to test the scale evolution predicted by pQCD)
a much better statistical accuracy and a significantly larger coverage in $p_T$ are needed. 
As a rule of thumb, a measurement of
a final state jet or hadron with a certain $p_T$ predominantly probes the
gluon density at momentum fractions $x\gtrsim 2p_T/\sqrt{S}$ where
$\sqrt{S}$ is the available proton-proton center-of-mass system (c.m.s.) energy.
Since the cross section drops rapidly with increasing $p_T$ it will take
RHIC quite some time to accumulate enough events to determine $\Delta g$ at
$x\gtrsim 0.1$. This is where the fixed-target experiments at
much lower energies can 
possibly help to add information about $\Delta g$ at medium-to-large $x$
in the very near future.

In the case of hadron-pair production, $l N\rightarrow l'H_1 H_2 X$, the process already
considered by {\sc Compass} and {\sc Hermes} \cite{ref:twohadron}, 
the theoretical framework is rather complex and so
far NLO corrections are still lacking. In this paper we therefore
focus on the {\em photoproduction} of {\em single-inclusive} hadrons,
$l N\rightarrow l'HX$, where the complete
NLO pQCD framework is already at hand \cite{ref:pionnlo,ref:nlophoton}
and has been recently applied to study the physics case of
a conceivable first polarized $ep$-collider at BNL \cite{ref:erhic}. 
Photoproduction with quasi-real photons has the
advantage of much higher rates than deeply-inelastic 
electroproduction of hadrons. 
The price to pay is the more involved theoretical framework for
photoproduction where so-called ``direct'' and ``resolved'' photons
contribute to the cross section \cite{ref:klasen} as depicted 
in Figs.~\ref{fig:cartoon} (a) and (b), respectively. 
Estimates for the latter require knowledge of the parton
content of circularly polarized photons which is lacking at the moment.
We will demonstrate below that this does not, however, seriously limit the 
usefulness of this process. 

We note that there are actually already some photoproduction data for 
double-longitudinal spin asymmetries in $e N\rightarrow H^{\pm} X$
from the {\sc E155} experiment at SLAC~\cite{ref:e155}.
Here the scattered electron was not observed, so the data
set is integrated over all virtualities of the exchanged photon.
Nonetheless, thanks to the photon propagator, production by
almost real photons strongly dominates.
The {\sc E155} data have been compared to the theoretical predictions 
of~\cite{ref:afanasev}, which were based on LO calculations
supplemented by certain higher-twist contributions. A relatively 
poor overall agreement was found. To achieve a better description
of the data, the  theoretical framework used in~\cite{ref:afanasev}
was subsequently augmented by estimates~\cite{ref:afanasev1} of soft 
contributions that are mostly relevant at small transverse momenta.
We will later present a brief comparison of our NLO calculations
with the {\sc E155} data. 

At fixed-target energies it is much less clear that pQCD methods are
applicable than at collider energies. In fact, data for high-$p_T$ 
processes in hadron-hadron fixed-target scattering have in the past
been a serious challenge for the factorized framework outlined 
above~\cite{ref:fixed-target}. It is therefore even more
important to demonstrate first that unpolarized production rates can
be described with standard pQCD methods before turning to studies of
spin asymmetries. Otherwise conclusions about the parton and/or spin content
of nucleons might be misleading or wrong. We will therefore not
only present results for spin asymmetries and discuss their possible sensitivity to
$\Delta g$ but also focus on predictions for unpolarized reference or
``benchmark'' cross sections which allow to probe the pQCD framework.

The paper is organized as follows: in Sec.~\ref{sec2} we briefly recall the
theoretical framework for photoproduction required for the subsequent
numerical sections and set up our notation.
We refrain from repeating lengthy technical details and refer the
interested reader to Refs.~\cite{ref:pionnlo,ref:nlophoton}.
We then turn to a  detailed study of single-inclusive pion
production in the kinematical regions relevant for {\sc Compass} and {\sc Hermes}
including, as far as possible, experimental cuts.
Detailed phenomenological results are presented in Secs.~\ref{sec3} and
\ref{sec4} for {\sc Compass} and {\sc Hermes}, respectively. Special emphasis is put
on theoretical uncertainties associated with the application
of perturbative methods in the domain of the rather moderate 
transverse momenta
accessible in fixed-target experiments. We first present unpolarized ``benchmark''
cross sections for each experiment and then turn to a discussion of 
the relevant spin asymmetries and their sensitivity to $\Delta g$.
We comment on the relevance of resolved photon contributions and their
influence on the extraction of the gluon polarization. In Sec.~\ref{sec3e155}
we present the comparison to the {\sc E155} data. We conclude in Sec.~\ref{sec5}.

%
\section{Technical Prerequisites}\label{sec2}
%
In the following, we consider the spin-dependent photoproduction cross section
$lN \to l^{\prime} H X$ for the scattering of a longitudinally polarized lepton beam $l$ off a 
target nucleon $N$ with subsequent hadronization of a produced parton $c$
into the observed hadron $H$ with momentum $P_H$, see Fig.~\ref{fig:cartoon}.
The observed hadron $H$ is at high transverse momentum $p_T$, ensuring the
required large momentum transfer. 
The differential single-inclusive cross section can then be schematically written as
a convolution 
\begin{eqnarray}
\label{eq:xsecdef}
d\Delta\sigma &\equiv&
\frac{1}{2}\left[ d\sigma_{++} - d\sigma_{+-} \right] \\ 
\label{eq:photoprod-xs}
&&\!\!\!\!\!\!\!\!\!\!\!\!\!\!\!\!\!\!\!= \sum_{a,b,c}
        \int dx_a \, dx_b \, dz_c 
        \,\Delta f^l_a(x_a,\mu_f)\,\Delta f^N_b(x_b,\mu_f)
        \non\\
&&\!\!\!\!\!\!\!\!\!\!\!\!\!\!\!\!\!\!\!\times
        d\Delta\hat{\sigma}_{ab\to cX}
        (S,x_a,x_b,P_H/z_c,\mu_r,\mu_f,\mu_f')
         D_c^H(z_c,\mu_f')\;.\nonumber\\
&&
\end{eqnarray}     
The subscripts ``$++$'' and ``$+-$'' in Eq.~(\ref{eq:xsecdef}) 
denote the helicities of the colliding leptons and nucleons.

In Eq.~(\ref{eq:photoprod-xs}) $x_b$ is the fraction of the 
nucleon momentum taken by parton $b$, and the $\Delta f_b^N(x_b,\mu_f)$ are the 
usual polarized parton densities. For instance,
the spin-dependent gluon distribution at a scale $\mu_f$ is defined by
\begin{equation}
\label{eq:pdf-def}
\Delta g\,(x_b,\mu_f)\equiv g_+(x_b,\mu_f)-g_-(x_b,\mu_f)\;,
\end{equation}
where the subscripts $\pm$ indicate the helicity of a gluon in a nucleon $N$
of positive helicity. 
The non-perturbative functions $D_c^H(z_c,\mu_f')$ describe the
fragmentation of parton $c$  ($c=q,\bar{q},g$) into the observed
hadron $H$ at a momentum scale $\mu_f'$. $z_c$ is the fraction of
parton $c$'s momentum taken by the hadron $H$.
The sum in Eq.~(\ref{eq:photoprod-xs}) runs over all partonic channels $a+b\to c+X$ 
contributing to the single-inclusive cross section $lN \to l^{\prime} H X$
with $d\Delta\hat{\sigma}_{ab\to cX}$ the associated spin-dependent
partonic hard scattering cross sections. The latter can be calculated in pQCD
order-by-order in the strong coupling constant $\alpha_s(\mu_r)$,
with $\mu_r$ denoting the renormalization scale, and are known up to
NLO accuracy. The spin-dependent results can be found in
\cite{ref:pionnlo,ref:nlophoton,ref:nlopoldir}.
The experimentally measured cross section is the {\em sum} of 
the so-called ``direct'' and ``resolved'' photon contributions, cf.\
Figs.~\ref{fig:cartoon} (a) and (b), respectively,
\begin{equation}\label{eq:xs-tot}
d\Delta\sigma = d\Delta\sigma_{\mathrm{dir}} + d\Delta\sigma_{\mathrm{res}}\;,
\end{equation}
which both can be cast into the form of Eqs.~(\ref{eq:xsecdef}) and (\ref{eq:photoprod-xs})
by defining $\Delta f_a^l$ appropriately.
Most generally, the parton densities of the lepton $l$, $\Delta f_a^l$,
can be written as convolutions,
\begin{equation}
\label{eq:convol}
\Delta f_a(x_a,\mu_f)=\int_{x_a}^1 \frac{dy}{y}\,\Delta P_{\gamma l}(y)
         \,\Delta f_a^\gamma\left(x_\gamma=\frac{x_a}{y},\mu_f\right),
\end{equation}
with 
\begin{eqnarray}
\label{eq:weiz-will}
\nonumber
\Delta P_{\gamma l}(y)&=&\frac{\alpha_e}{2\pi}\Bigg\{
                \left[\frac{1-(1-y)^2}{y}\right]\ln\frac{
                Q_{\max}^2(1-y)}{m_l^2 y^2}\\
&+& 2m_l^2 y^2\left(\frac{1}{Q_{\max}^2}-\frac{1-y}{m_l^2 y^2}\right)\Bigg\}
\end{eqnarray}
being the spin-dependent Weizs\"acker-Williams ``equivalent photon'' spectrum
\cite{ref:daniel} that describes the collinear emission of a photon with low virtuality 
$Q^2$ less than some upper limit $Q_{\max}^2$ (determined by the experimental conditions)
by a lepton of mass $m_l$.

The explicit form of the $\Delta f^\gamma_a$ in Eq.~(\ref{eq:convol}) depends
on the specifics of the interaction that the quasi-real photon radiated off the
lepton undergoes in the hard scattering.
For the {\em direct} photon contribution to the cross section, $d\Delta\sigma_{\mathrm{dir}}$,
depicted in Fig.~\ref{fig:cartoon} (a), parton $a$ in Eq.~(\ref{eq:photoprod-xs})
has to be identified with an elementary photon and hence $x_a$ 
with the momentum fraction $y$ of the photon w.r.t.\ the parent lepton, i.e.,
\begin{equation}
\Delta f_a^\gamma = \delta(1-x_\gamma)
\end{equation}
in Eq.~(\ref{eq:convol}).
If the photon first resolves into hadronic ``constituents'' which then 
interact with partons of the target nucleon $N$, $\Delta f_a^\gamma$ in 
Eq.~(\ref{eq:convol}) represents the parton densities of a circularly polarized
photon. The latter are defined in complete analogy to the ones for a nucleon 
target in Eq.~(\ref{eq:pdf-def}). 
Unlike hadronic parton distributions which are genuinely non-perturbative
objects, photonic densities consist of a perturbatively calculable ``pointlike''
contribution which dominates their behavior at large momentum fractions
$x_\gamma$ in Eq.~(\ref{eq:convol}). This will become important in the
discussion of the numerical results in the remainder of the paper.

We shall note that neither $d\Delta\sigma_{\mathrm{dir}}$ nor $d\Delta\sigma_{\mathrm{res}}$ individually
are measurable cross sections as their definition depends on the specific
details of the factorization procedure for dealing with singular collinear
parton emissions. Only their sum, Eq.~(\ref{eq:xs-tot}), is a physically
meaningful quantity provided both $d\Delta\sigma_{\mathrm{dir}}$ and 
$d\Delta\sigma_{\mathrm{res}}$ have been calculated consistently in the same 
factorization scheme. Needless to say, the corresponding spin-averaged cross sections are 
straightforwardly obtained by replacing all polarized parton densities,
partonic cross sections, etc., by their appropriate unpolarized counterparts.

Finally, we wish to stress that Eqs.~(\ref{eq:xsecdef}) and
(\ref{eq:photoprod-xs}) apply to a single-inclusive cross section, which 
is not to be confused with a cross section for producing a 
so-called {\em ``leading hadron''}. The latter, for which 
only the hadron with the highest momentum in the event
is counted, is sometimes provided by experiment.
It is not possible in the framework of pQCD based on the
factorization theorem and outlined above to compute a ``leading-hadron''
cross section, as fragmentation functions $D_c^H$ are fully inclusive.

\section{Phenomenological Results}\label{sec30}
%
With the technical framework and notation set up
we are now in a position to perform a detailed phenomenological analysis of
polarized and unpolarized single-inclusive pion photoproduction cross sections 
for the {\sc Compass} and {\sc Hermes} fixed-target experiments in Secs.~\ref{sec3} 
and \ref{sec4}, respectively.
In Sec.~3.3 we present the comparison to the {\sc E155} data.

All our results will be differential in the hadron's transverse
momentum $p_T$ and integrated over all kinematically and experimentally 
allowed pseudo-rapidities $\eta$ of the produced hadron $H$ 
measured w.r.t.\ the direction of the incident lepton beam.
For the unpolarized parton densities of the nucleon
and photon we adopt throughout the CTEQ \cite{ref:cteq} and GRV
\cite{ref:grvphoton} sets, respectively. 
To study the sensitivity to the
unknown gluon polarization of the nucleon we use four different sets of
spin-dependent parton distributions emerging from the GRSV analysis \cite{ref:grsv}.
These sets span a rather large range of gluon densities $\Delta g$ 
all consistent with present DIS data. Apart from the ``standard'' set of
GRSV with a moderately large, positive $\Delta g$, the three other sets
``$\Delta g=g$ input'', ``$\Delta g=0$ input'', and ``$\Delta g=-g$ input'' 
are characterized by a large positive, a vanishing, and a large negative gluon
polarization, respectively, at the input scale of the evolution.

To model the hadronization into pions, we use the fragmentation functions of KKP
\cite{ref:kkp} which are known to provide a good description of the
$pp\rightarrow \pi^0 X$ cross sections measured by {\sc Phenix} and {\sc Star}
\cite{ref:rhic-unppion} (and, of course, all available $e^+e^-$ data).
Although the framework outlined in Sec.~\ref{sec2} applies in principle to the
photoproduction of {\em any} hadron species (made of light quark flavors only),
we limit ourselves mainly to the production of neutral pions. These
are most abundant, and the $\pi^0$ fragmentation
functions, mainly extracted at large resolution scales 
$\mu\simeq M_Z$ from LEP data, have been shown to work quite well also at 
scales of a few GeV relevant here \cite{ref:fraglow}.

Since the parton distributions of circularly polarized photons,
$\Delta f^\gamma$, required for estimates of the resolved photon
contribution $d\Delta\sigma_{\mathrm{res}}$, are completely
unknown so far one has to resort to some models.
For our purposes it is sufficient to use the two extreme scenarios
proposed in Ref.~\cite{ref:photmodels} which are
based on ``maximal'' $[\Delta f^\gamma(x,\mu_0) = f^\gamma(x,\mu_0)]$ and ``minimal'' 
$[\Delta f^\gamma(x,\mu_0) = 0]$ saturation of the positivity condition 
$|\Delta f^\gamma(x,\mu_0)| \leq f^\gamma(x,\mu_0)$ at the starting scale
$\mu_0$ for the evolution to $\mu>\mu_0$. 
Both models result in very different parton distributions $\Delta f^\gamma$ at
small-to-medium $x_\gamma$ while they almost coincide as $x_\gamma\to 1$
due to the dominance of the perturbatively calculable pointlike contribution
in this region. Unless we study the impact of the details of the
non-perturbative input to the evolution of $\Delta f^\gamma$ on the
the full photoproduction cross section, 
the use of  the ``maximal'' set will be implicitly understood. 
For recent work on the parton distributions of polarized 
photons, see~\cite{ref:grs}.

\subsection{Single-Inclusive Pion Production at COMPASS}\label{sec3}
%
With the present setup, {\sc Compass} scatters polarized muons
off the deuterons in a polarized $^6$LiD solid-state target. The 
beam energy is $E_\mu=160$~GeV, corresponding to a lepton-nucleon 
c.m.s.\ energy of $\sqrt{S}\simeq 18$~GeV.
On average the beam polarization is ${\cal{P}}_\mu\simeq 76\%$.
About ${\cal{F}}_d\simeq50\%$ (``dilution factor'') of the deuterons can be polarized,
with an average polarization of ${\cal{P}}_d\simeq 50\%$ \cite{ref:compass}.
It is also conceivable to run in the future with a ``proton target'' (NH$_3$)
which, however, will have a much less advantageous dilution factor of
about ${\cal{F}}_p\simeq 17.6\%$ but a polarization of ${\cal{P}}_p\simeq 85\%$.
We will therefore mainly present results for the photoproduction of
neutral pions off a deuteron target. 

%
\begin{figure}[ht]
\vspace*{-0.75cm}
\begin{center}
\includegraphics[width=0.49\textwidth,clip=]{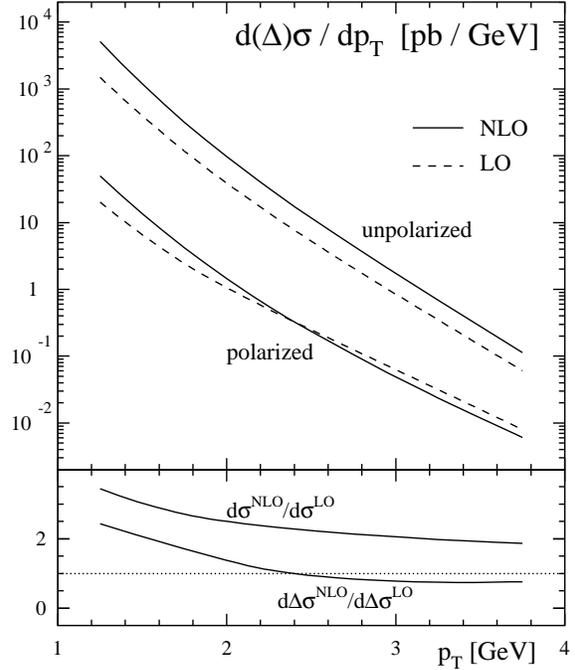}     
\vspace*{-0.4cm}
        \caption{Unpolarized and polarized $p_T$-differential
                single-inclusive cross sections at LO (dashed) and NLO (solid) for
                the photoproduction of neutral pions, $\mu d\to \mu'\pi^0 X$ at 
                $\sqrt{S}=18$~GeV, integrated over the angular acceptance of
                {\sc Compass}. 
                The lower panel shows the ratios of NLO to LO contributions
                ($K$-factor).} \label{fig:comp-xs-kfac}        
\end{center}
\end{figure}
%
%
Currently, pions can be detected if
their scattering angle is less than $\theta_{\max}=70$~mrad. Using
$\eta=-\ln\tan\left(\theta/2\right)$, this 
straightforwardly translates into a minimal bound on pseudo-rapidity
$\eta^{\min}\simeq 3.35$ in the laboratory frame, corresponding
to $\eta_{cms}^{\min}\simeq 0.44$ in the lepton-nucleon c.m.s.\
where we have made use of the well-known behavior of 
rapidity under Lorentz boosts:

\begin{equation}
\label{eq:eta-boost}
\eta_{cms} = \eta - \frac{1}{2}\ln\frac{2 E_\mu}{M_N}\;,
\end{equation}
with $M_N$ the nucleon mass.
Kinematics sets an additional upper bound on the pion's rapidity
depending on its transverse momentum: 
$\eta_{cms}^{\max}=\cosh^{-1}({\sqrt{S}/2 p_T})$.
As already mentioned, 
we will always integrate over all kinematically allowed rapidities,
$0.44\lesssim\eta_{cms}\lesssim \cosh^{-1}({\sqrt{S}/2 p_T})$. 

The momentum distribution of the quasi-real photons radiated off the
muons can be described by the Weiz\-s\"acker-Williams spectrum given in
Eq.~(\ref{eq:weiz-will}) with $m_l=m_\mu$ and $Q^2_{\max}=0.5$~GeV$^2$. 
The photon's momentum fraction $y$ is restricted to be in the range
$0.2\leq y\leq 0.9$. At smaller $y$ the photon polarization is strongly
diluted as the unpolarized equivalent photon spectrum behaves like
$(1+(1-y)^2)/y$ rather than $(1-(1-y)^2)/y$ in Eq.~(\ref{eq:weiz-will}).
We note that the often omitted non-logarithmic pieces in Eq.~(\ref{eq:weiz-will})
result in a small but non-negligible contribution here.

%
\begin{figure}[ht]
\vspace*{-0.75cm}
\begin{center}
\includegraphics[width=0.49\textwidth,clip=]{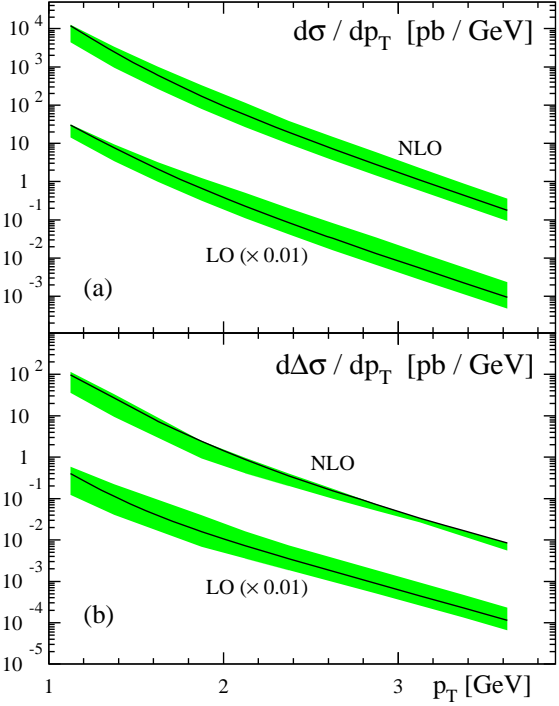}  
\vspace*{-0.4cm}    
        \caption{Scale dependence of the LO and NLO unpolarized {\bf (a)}
                 and polarized {\bf (b)} cross sections 
                 for $\mu d\to \mu' \pi^0 X$ shown in Fig.~\ref{fig:comp-xs-kfac}. 
                 All scales are varied simultaneously in the range 
                 $p_T/2\leq\mu_r=\mu_f=\mu_f'\leq 2 p_T$. 
                 Solid lines correspond to the choice where all scales are 
                 set to $p_T$. 
                 All LO computations have been rescaled by a factor 0.1
                 to better distinguish them from the NLO results. 
              } \label{fig:compass-scdep}          
\end{center}
\end{figure}
%
%
Figure~\ref{fig:comp-xs-kfac} shows our predictions for the 
$p_T$-differential cross section for the polarized and
unpolarized photoproduction of neutral pions,  $\mu d\to \mu' \pi^0 X$,
at LO and NLO accuracy at {\sc Compass}.
We have set all factorization and renormalization scales in 
Eq.~(\ref{eq:photoprod-xs}) equal to the pion transverse momentum, 
$\mu_f=\mu_f'=\mu_r=p_T$. 
The so-called $K$-factor, defined as the ratio of NLO to LO 
unpolarized (polarized) cross sections,
\begin{equation}
\label{eq:k-factor}
K=\frac{d(\Delta)\sigma^{\mathrm{NLO}}}{d(\Delta)\sigma^{\mathrm{LO}}}\;,
\end{equation}
is depicted in the lower panel of Fig.~\ref{fig:comp-xs-kfac}. 
The results indicate clearly the relevance of the NLO corrections to
the cross section in the small-to-medium $p_T$-region accessible at
a fixed-target experiment, in particular in the unpolarized case.
The effects of higher order corrections are less pronounced 
for the longitudinally polarized cross section, mainly due to 
large cancellations among the various
partonic channels contributing to $d\Delta\sigma$ at NLO.
The different behavior of the unpolarized and polarized $K$-factors also clearly indicates that
the contributions due to NLO corrections do {\em not} cancel in the experimentally relevant double-spin asymmetry
defined as
\begin{equation}
\label{eq:spinasy-def}
A_{\mathrm{LL}}^{H,N}\equiv\frac{d\Delta\sigma}{d\sigma}=
        \frac{d\sigma_{++}-d\sigma_{+-}}{d\sigma_{++}+d\sigma_{+-}}\;.
\end{equation}

Not unexpectedly, at $p_T\simeq 1\div 2 \,\mathrm{GeV}$, the $K$-factors for both, polarized and
unpolarized cross sections rise sharply, perhaps indicating a breakdown of the
standard pQCD framework as outlined in Sec.~\ref{sec2}.
Since,
as will be demonstrated below,  
this is precisely the $p_T$-region where the statistical accuracy
of {\sc Compass} would best allow to deduce some information 
about $\Delta g$ from a measurement of
the double-spin asymmetry $A_{\mathrm{LL}}^{\pi,d}$,
one has to ensure the validity of the pQCD
framework first. As already emphasized in the Introduction, this is best achieved
by a measurement of the unpolarized cross section shown in Fig.~\ref{fig:comp-xs-kfac},
where all ingredients, partonic cross sections, parton distributions, and 
fragmentation functions are known.
We note that all-order resummations of large logarithms
in the perturbative series which appear when the initial partons 
have just enough energy to produce a high-$p_T$ pion and a recoiling massless ``jet''
may lead to a considerable enhancement of the cross section 
at fixed-target energies as was recently demonstrated for the process $pp\rightarrow \pi X$ 
\cite{ref:resum}. Similar calculations for the case of photoproduction
are not yet available but certainly desirable. 
Any residual shortfall of the resummed theoretical prediction would then 
indicate the relevance of non-perturbative contributions.

Large $K$-factors, as found in Fig.~\ref{fig:comp-xs-kfac}, are, however, of limited
significance for unambiguously estimating 
the impact of higher order corrections in a perturbative calculation.
This is due to the large scale uncertainties associated with the LO cross
sections entering the denominator of Eq.~(\ref{eq:k-factor}).
We therefore further explore the reliability of the perturbative
approach by studying the dependence of the calculated cross sections,
Eq.~(\ref{eq:photoprod-xs}), on the unphysical, a priori arbitrary factorization and
renormalization scales, $\mu_f$, $\mu^\prime_f$ and $\mu_r$, respectively. 
Any dependence on these scales is a remnant of the truncation
of the perturbation series at some fixed order of $\alpha_s$ and thus
expected to diminish if higher order corrections are included. This is the
prime motivation for going beyond the LO approximation of pQCD. 
The scales are of the order of the hard scale characterizing the process,
here, the large $p_T$ of the observed hadron, but not further specified by theory. 
An estimate for the sensitivity of the computed cross section to $\mu_f$,
$\mu^\prime_f$, and $\mu_r$ is usually obtained by varying them collectively in the range
$p_T/2\leq \mu_f=\mu_f'=\mu_r\leq 2 p_T$. We note that in principle all scales
can be varied independently.  

%
%
\begin{figure}[htp]
\vspace*{-0.75cm}
\begin{center}
\includegraphics[width=0.49\textwidth,clip=]{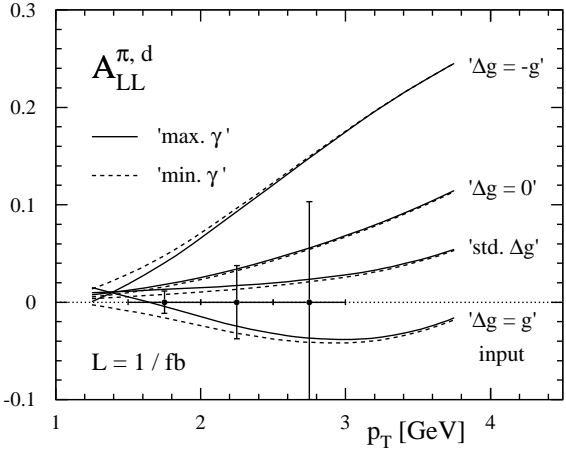}     
        \caption{Double-spin asymmetry $A_{\mathrm{LL}}^{\pi,d}$ at NLO
        for different gluon polarizations in the nucleon and 
        minimal and maximal saturation of the polarized photon densities, 
        dashed and solid lines, respectively (see text). 
        The ``error bars'' indicate the estimated statistical
        uncertainty for such a measurement at {\sc Compass}.  
              } \label{fig:comp-asy} 
\end{center}
\end{figure}
%
%
The shaded bands in Figs.~\ref{fig:compass-scdep} (a) and (b)
indicate the resulting scale uncertainty of the unpolarized and polarized cross
sections, respectively, shown in Fig.~\ref{fig:comp-xs-kfac}.
In contrast to similar studies for inclusive high-$p_T$ pion hadro- \cite{ref:pionnlo}
and photoproduction \cite{ref:nlophoton} at collider energies, 
where the theoretical scale uncertainties are substantially reduced when the
NLO corrections are taken into account, this barely happens here.
This is particularly true for the unpolarized cross section, whereas the scale
dependence of the polarized cross section improves beyond the LO, but only
slightly in the region $1\le p_T \le 2\,\mathrm{GeV}$ which mainly matters for
a determination of $\Delta g$.
Together with the large $K$-factors found in Fig.~\ref{fig:comp-xs-kfac}
this underlines the delicacy of a perturbative calculation in the low-energy range
associated with fixed-target experiments such as {\sc Compass}. 
It is therefore particularly important 
to check the applicability of pQCD methods 
by showing, for instance, that data taken in
unpolarized collisions fall within the uncertainty band shown in Fig.~\ref{fig:compass-scdep}.

Next we consider the double spin asymmetry $A_{\mathrm{LL}}^{\pi,d}$,
Eq.~(\ref{eq:spinasy-def}), for single-inclusive neutral pion
production which will be one of the main quantities of interest
in experiment. Fig.~\ref{fig:comp-asy} shows $A_{\mathrm{LL}}^{\pi,d}$,
calculated at NLO for the ``standard'' set of the GRSV spin-dependent
parton densities \cite{ref:grsv}, as well as for three other sets 
emerging from the GRSV analysis which mainly differ in the 
assumptions about $\Delta g$ (see above).
The impact of the unknown non-perturbative parton structure of the circularly
polarized photon on $A_{\mathrm{LL}}^{\pi,d}$ is examined by making use of the two extreme 
sets also introduced at the beginning of Sec.~\ref{sec30}.
We refrain from showing LO estimates for the double-spin asymmetry which are
of rather limited use anyway. Due to the pronounced differences in the
$K$-factors for the unpolarized and polarized cross sections, see
Fig.~\ref{fig:comp-xs-kfac}, the LO results for $A_{\mathrm{LL}}^{\pi,d}$ are 
considerably larger than the NLO ones shown in Fig.~\ref{fig:comp-asy}.   
This is in contrast to the frequently made assumption that NLO
corrections cancel in spin asymmetries. We note that similar observations
have been also made for hadroproduction of pions \cite{ref:pionnlo}.

As can be seen, the actual choice
of photonic parton densities barely affects the results for the
spin asymmetry shown in Fig.~\ref{fig:comp-asy} if the pion's
transverse momentum is larger than about $2\,\mathrm{GeV}$.
This can be readily understood by noticing that in this region 
the average momentum fraction $\langle x_a \rangle$ in Eq.~(\ref{eq:convol}) 
is larger than 0.5, i.e., one probes only $x_{\gamma}$-values 
where the photon structure is dominated by the ``pointlike'' contribution
independent of the details of the unknown non-perturbative input
\cite{ref:photmodels}.
In this $p_T$-region a measurement of the spin asymmetry could be
related to a certain $\Delta g$. However, this does not imply that
the resolved contribution to the cross section is negligible as we shall
demonstrate below. 
On the other hand, for $p_T\lesssim 1.5\,\mathrm{GeV}$ results for different 
gluon polarizations of the nucleon and assumptions about $\Delta f^{\gamma}$ 
overlap, making it virtually impossible to draw any conclusions without either
knowing $\Delta g$ or $\Delta f^{\gamma}$.

To judge whether the observed dependence of $A_{\mathrm{LL}}^{\pi,d}$ on $\Delta g$ 
in Fig.~\ref{fig:comp-asy} can be used to learn something about the gluon polarization, 
we give estimates for the expected statistical accuracy $\delta A_{\mathrm{LL}}^{\pi,d}$ 
for such a measurement at {\sc Compass} 
in certain bins of $p_T$, calculated from
\begin{equation}
\label{eq:all-error}
\delta A_{\mathrm{LL}}^{H,N} \simeq \frac{1}{{\cal{P}}_\mu {\cal{P}}_N {\cal{F}}_N} 
\frac{1}{\sqrt{\sigma_{bin}\mathcal{L}}}\;\,.
\end{equation}
Here, $\sigma_{bin}$ denotes the unpolarized cross section in the $p_T$-bin
considered, ${\mathcal{L}}$ the integrated luminosity for which we assume 
$1\,\mathrm{fb}^{-1}$, and all other parameters are as specified at 
the beginning of Sec.~\ref{sec3}.
Not unexpectedly, the statistical accuracy rapidly deteriorates towards
higher $p_T$ as the photoproduction cross section drops sharply.
In the region $1.5 \lesssim p_T \lesssim 2.5\,\mathrm{GeV}$ it is conceivable
that some information about $\Delta g$ can be obtained. We note that the 
gluon polarization will be predominantly probed in the range 
$0.1\lesssim x_b\lesssim 0.3$
of the momentum fraction $x_b$ in Eq.~(\ref{eq:photoprod-xs}).

\begin{figure}[thp]
\vspace*{-0.75cm}
\begin{center}
\includegraphics[width=0.495\textwidth,clip=]{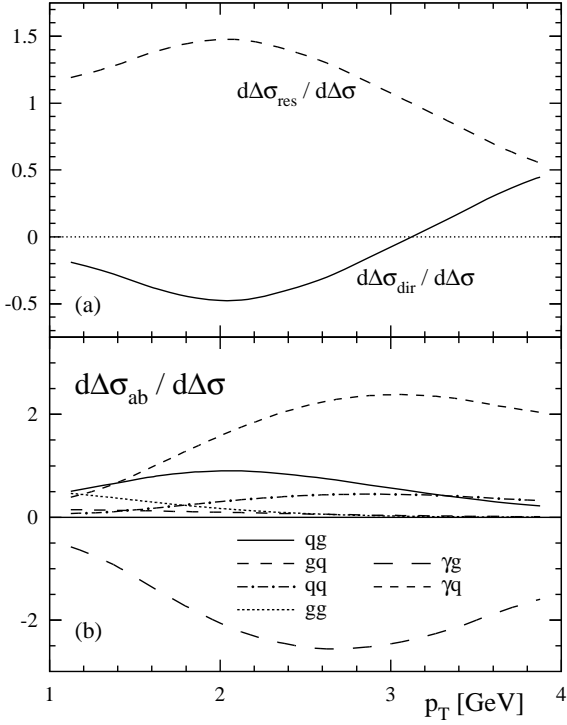}     
        \caption{{\bf (a)} Direct and resolved NLO($\overline{\mathrm{MS}}$) contributions and
          {\bf (b)} NLO($\overline{\mathrm{MS}}$) contributions of different 
            partonic channels $a+b \to c+X$ 
            to the full NLO polarized photoproduction cross section
            shown in Fig.~\ref{fig:comp-xs-kfac}.}\label{fig-comp-subproc}
\end{center}
\end{figure}

Let us now turn to a closer analysis of how the results presented in
Figs.~\ref{fig:comp-xs-kfac} and \ref{fig:comp-asy} come about and can be
understood, by studying the different contributions to the polarized photoproduction 
cross section separately. This is done in Fig.~\ref{fig-comp-subproc}.
Here we use again our default sets of parton densities: the ``standard'' set
of GRSV \cite{ref:grsv} for the nucleon and the ``maximal'' set 
of \cite{ref:photmodels} for the photon.
First of all, Fig.~\ref{fig-comp-subproc}~(a) reveals that the resolved photon
cross section $d\Delta\sigma_{\mathrm{res}}$ dominates over the 
direct contribution $d\Delta\sigma_{\mathrm{dir}}$ in the entire range of $p_T$
relevant for {\sc Compass}. 
At a first glance this result is counter-intuitive and requires further
clarification since at fixed-target energies one might expect a dominance 
of the direct contributions to the full photoproduction cross section.
The importance of the resolved photon contribution, in particular in the polarized case, can be 
readily understood by a closer inspection of the individual 
partonic subprocesses contributing to the full photoproduction cross section.
%
%
%
%

Figure~\ref{fig-comp-subproc}~(b) shows that indeed both direct channels,
$\gamma q$ and $\gamma g$ scattering, exceed in magnitude
any of the resolved subprocesses. However, the two
channels contribute with different sign and therefore
cancel each other to a large extent in the sum $d\Delta \sigma_{\mathrm{dir}}$.
In this way numerically smaller resolved subprocesses can become important,
in particular, as they all contribute with the same sign for a positive
gluon polarization. 
We note that the partonic spin asymmetry $d\Delta\hat{\sigma}/d \hat{\sigma}$
for the LO process  $\gamma g\rightarrow q\bar{q}$ is $-1$, such that only positive
$\Delta g$ lead to this cancellation pattern. Nevertheless, also for
a large negative gluon polarization [``$\Delta g =-g$ input'' in
Fig.~\ref{fig-comp-subproc}~(b)] is the resolved contribution non-negligible. 
For all relevant values of $p_T$ the bulk of the resolved contribution stems from the
$qg$ subprocess where a quark at large momentum fraction $x_{\gamma}$
from the photon scatters off a gluon from the nucleon 
[solid line in Fig.~\ref{fig-comp-subproc}~(b)].
This is because quark densities in the photon are sizable at large
$x_{\gamma}$ \cite{ref:grvphoton,ref:photmodels}
due to the pointlike contribution, contrary to parton
densities in the nucleon which rapidly vanish as $x\rightarrow 1$.
As mentioned before, at large momentum fractions $\Delta q^\gamma$ is not 
sensitive to the details of how the non-perturbative hadronic content of the photon
was modeled. We therefore conclude that photoproduction of inclusive 
hadrons with {\sc Compass} kinematics and statistics will most likely not yield 
much information on the non-perturbative
nature of the resolved photon. 
On the other hand, this finding certainly simplifies an
analysis of photoproduction data with $p_T\gtrsim 1.5\,\mathrm{GeV}$
in terms of $\Delta g$ despite our ignorance of the photonic parton 
densities $\Delta f^{\gamma}$.
%
%
\begin{figure}[thp]
\begin{center}
\vspace*{-.5cm} 
\includegraphics[width=0.49\textwidth,clip=]{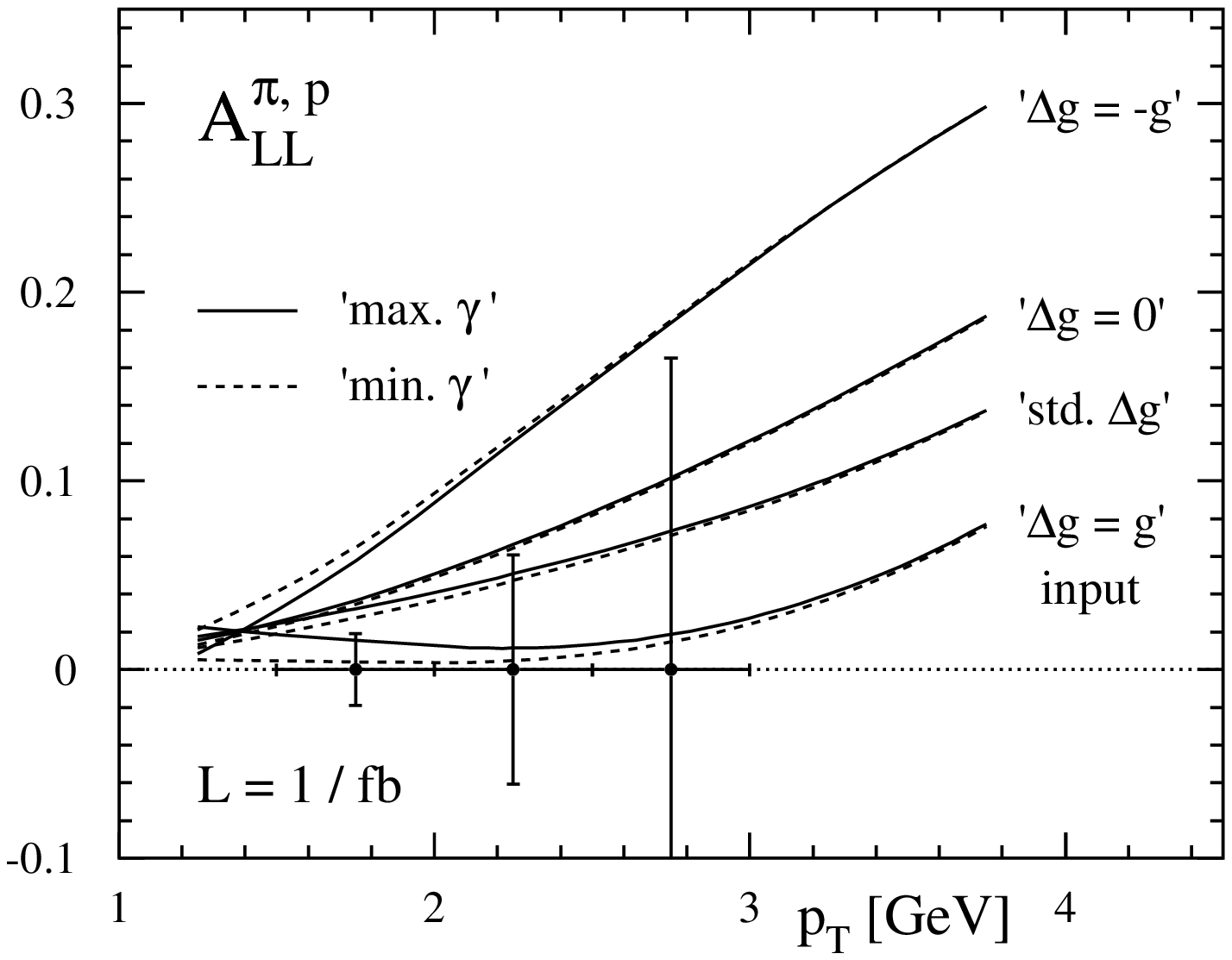}     
        \caption{As in Fig.~\ref{fig:comp-asy}, but now for a
        ``proton'' target.}\label{fig-comp-asy-proton}    
\includegraphics[width=0.49\textwidth,clip=]{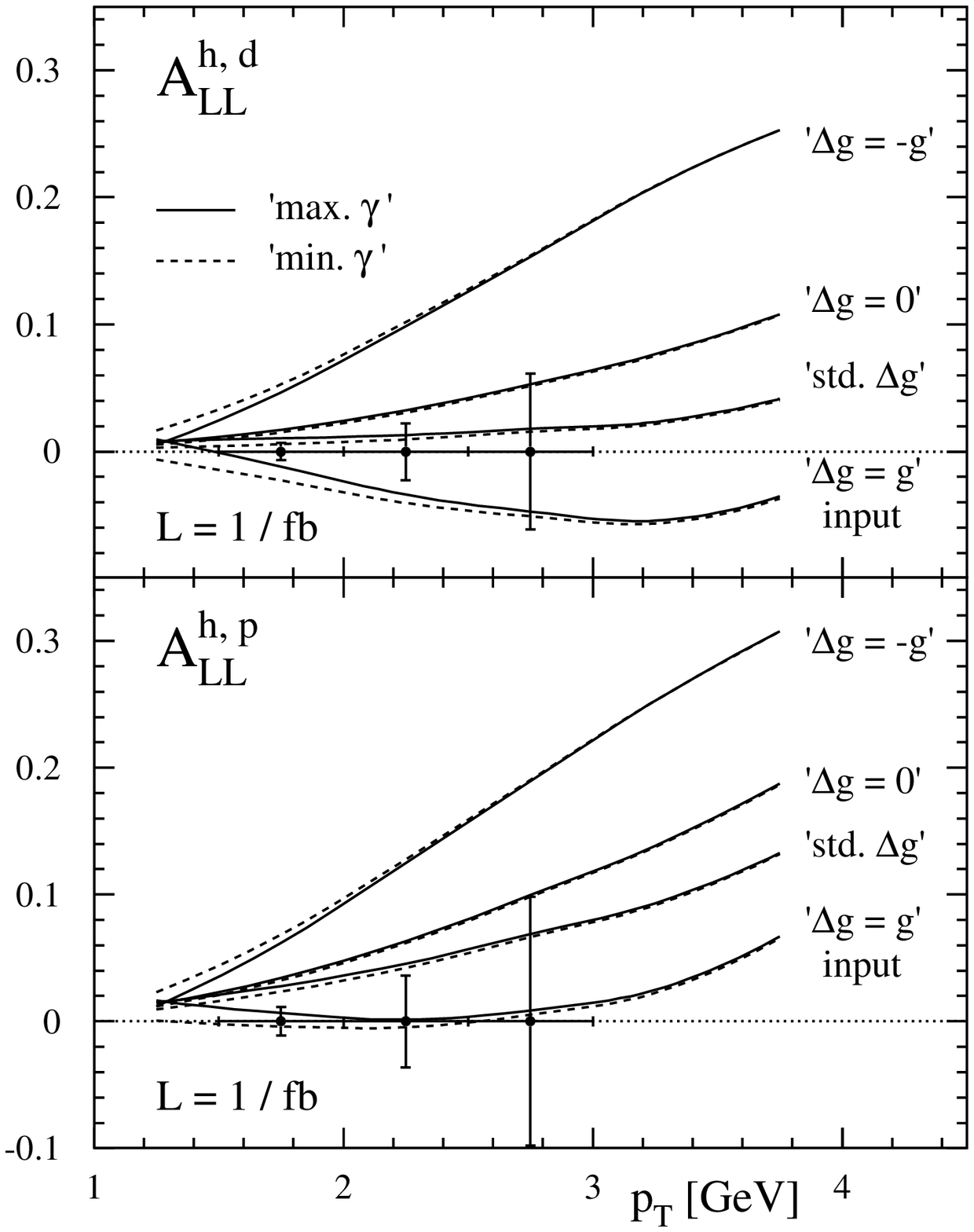}     
        \caption{As in Figs.~\ref{fig:comp-asy} and \ref{fig-comp-asy-proton},
         but now for the photoproduction of charged hadrons (see text).}
         \label{fig:comp-asy-chghad}

\end{center}
\end{figure}
%

We obtain very similar results when considering a proton target, 
as will be realized when the {\sc Compass} experiment switches to 
a $\mathrm{NH}_3$ target in the future. The resulting spin asymmetry, 
$A_{\mathrm{LL}}^{\pi,p}$, is depicted in Fig.~\ref{fig-comp-asy-proton}.
To estimate the statistical accuracy we have again assumed an integrated
luminosity of $1\,\mathrm{fb}^{-1}$, and all other parameters are as specified at 
the beginning of Sec.~\ref{sec3}.
The sensitivity to $\Delta g$ and the rather weak dependence on the photon
scenario for $p_T\gtrsim 1.5\,\mathrm{GeV}$, characteristic for the
large $x_\gamma$-region probed at {\sc Compass}, are essentially the same as
for $A_{\mathrm{LL}}^{\pi,d}$.
We note that the intricate interplay between direct and resolved photon processes
is quite similar to the one described in Figs.~\ref{fig-comp-subproc}~(a) and (b),
but that the cancellation between the $\gamma g$ and $\gamma q$ channels 
is less complete such that the resolved cross section is somewhat less
relevant here.

%
%
\begin{figure}[thp]
\begin{center}
\vspace*{-.55cm} 
\includegraphics[width=0.49\textwidth,clip=]{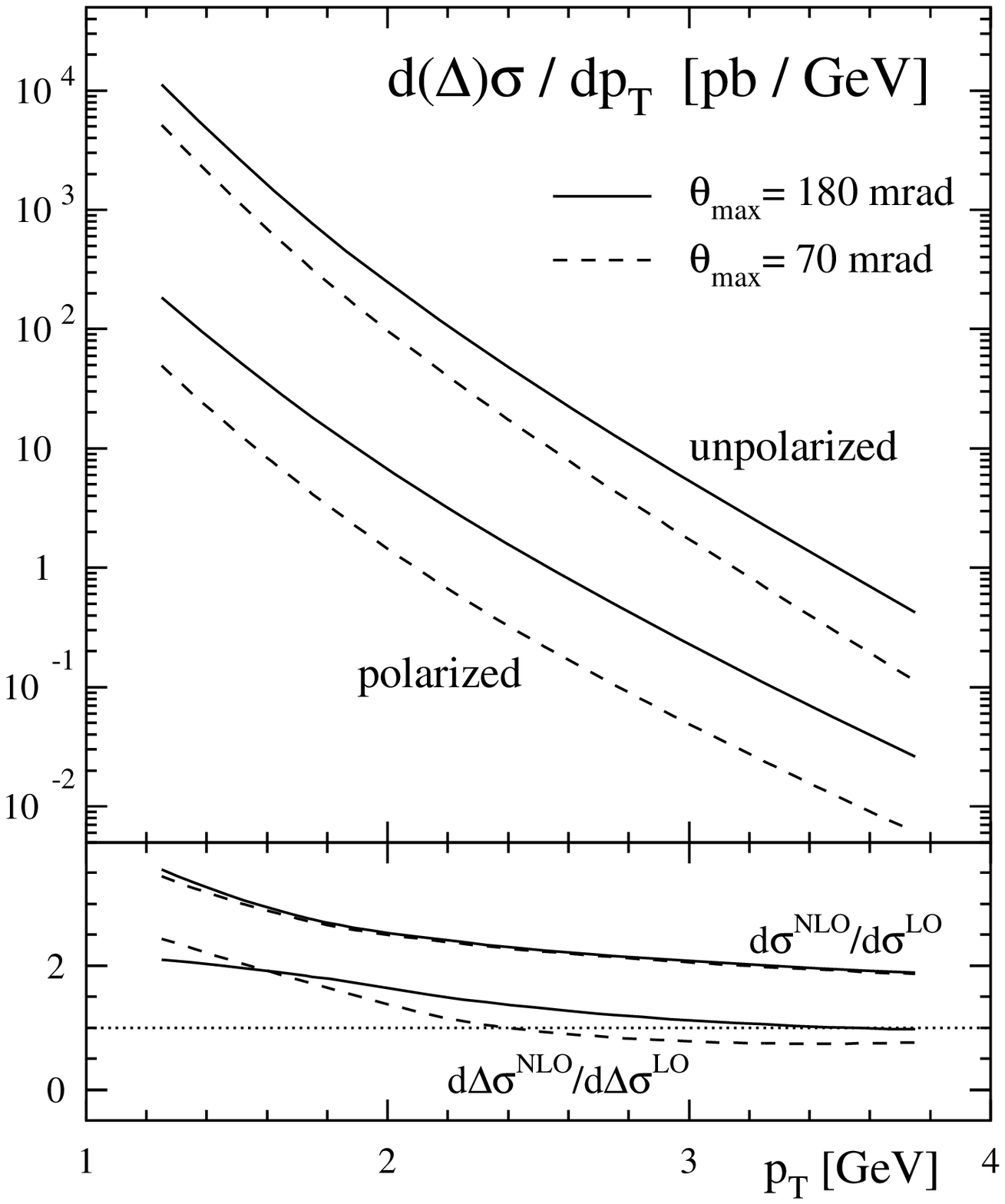}     
        \caption{Unpolarized and polarized $p_T$-differential
                cross sections at NLO for the 
                reaction $\mu d \to \mu' \pi^0 X$ for two different
                experimental setups: $\theta_{\max}=180$ mrad (solid) and,
                as in Fig.~\ref{fig:comp-xs-kfac},  
                $\theta_{\max}=70$ mrad (dashed).
                The lower panel shows the corresponding ratios of NLO and LO 
                results ($K$-factor).} \label{fig:comp-xsec-acceptance}  
\vspace*{-.3cm}
\includegraphics[width=0.49\textwidth,clip=]{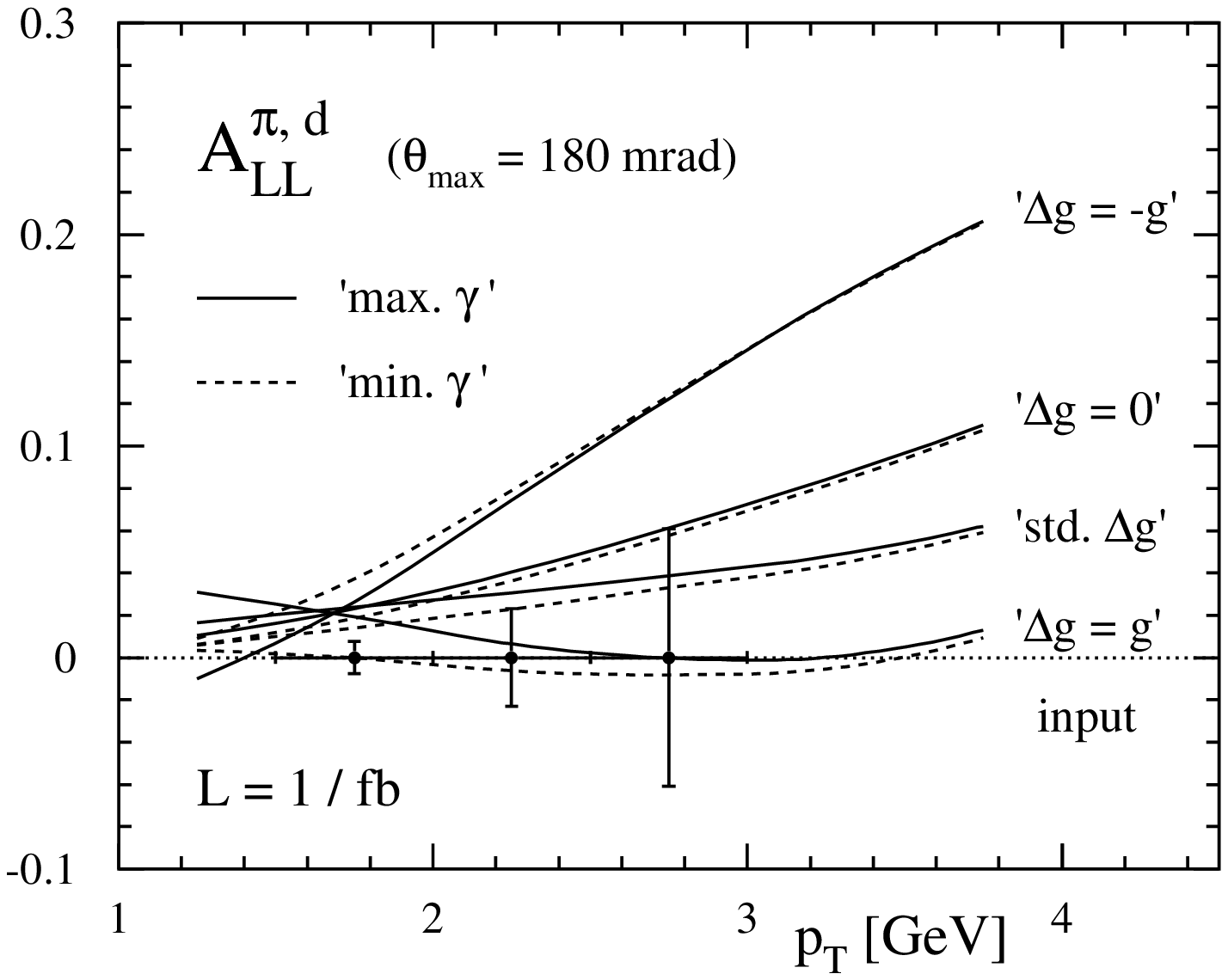}     
        \caption{As in Fig.~\ref{fig:comp-asy}, but now for $\theta_{\max}=180\,\mathrm{mrad}$. 
                }\label{fig:comp-asy-th}  
\end{center}
\end{figure}
%
%
So far we have only considered the production of neutral pions where fragmentation
functions were shown to work reasonably well also at rather 
low scales \cite{ref:fraglow}.
However, the sum of charged hadrons, predominantly pions, but also kaons and protons, is
equally important experimentally as these are often more easily identified than
neutral pions. This is also the case for {\sc Compass} at the moment.
In Fig.~\ref{fig:comp-asy-chghad} we therefore present the relevant spin asymmetries 
$A_{\mathrm{LL}}^{h,d}$ and $A_{\mathrm{LL}}^{h,p}$ for deuteron and proton targets, respectively,
for the reaction $\mu d(p) \to \mu' h X$, where $h$ represents the 
sum of charged hadrons (pions, kaons, and protons).
The results are obtained by employing the appropriate set of fragmentation 
functions, $D_c^{h^++h^-}$ of Ref.~\cite{ref:kkp}. 
Not unexpectedly, the gross features of the spin asymmetries in charged hadron production 
are the same as in neutral pion production. Due to the larger rate
for the sum of charged hadrons, the statistical precision is noticeably better
than for $A_{\mathrm{LL}}^{\pi,d}$ which makes such a measurement 
potentially more interesting.

From our results for the spin asymmetries shown in
Figs.~\ref{fig:comp-asy}, \ref{fig-comp-asy-proton}, and \ref{fig:comp-asy-chghad}
it is obvious that a major difficulty in extracting the gluon polarization
$\Delta g$ at fixed-target energies lies in the poor statistical accuracy
at large $p_T$-values. This, however, is the region where 
perturbation theory is expected to be more reliable and where the
uncertainties associated with the resolved photon contributions
to the cross section are much better under control.
It is therefore interesting to study whether the planned upgrade of the
{\sc Compass} experimental setup, which would lead to a much larger acceptance of
$\theta_{\max}=180\,\mathrm{mrad}$ and hence larger cross sections, 
could help.

In Fig.~\ref{fig:comp-xsec-acceptance} we compare the polarized and unpolarized 
neutral pion photoproduction cross sections 
for the present setup with $\theta_{\max}=70\,\mathrm{mrad}$ and for
the proposed upgrade with $\theta_{\max}=180\,\mathrm{mrad}$.
The gain in cross section is a factor $4\div5$ depending on the $p_T$-value considered,
yielding an improvement of the statistical accuracies for spin asymmetry
measurements by about a factor of two. The lower panel of Fig.~\ref{fig:comp-xsec-acceptance}
shows that the respective $K$-factors do not change significantly when going from
$\theta_{\max}=70\,\mathrm{mrad}$ to $\theta_{\max}=180\,\mathrm{mrad}$.
However, not only does an increased angular acceptance improve the statistical
accuracy, it also changes the interplay of direct and resolved subprocesses. 
It turns out that differences between the ``maximal'' and ``minimal'' $\Delta f^{\gamma}$
scenarios are more pronounced and persist towards larger $p_T$. This can be
inferred from our estimate for the spin asymmetry $A_{\mathrm{LL}}^{\pi,d}$ 
shown in Fig.~\ref{fig:comp-asy-th} if compared to the results given in
Fig.~\ref{fig:comp-asy}. Also the sensitivity of $A_{\mathrm{LL}}^{\pi,d}$ to different
assumptions about the gluon polarization of the nucleon is somewhat smaller
than before.

\subsection{Single-Inclusive Hadron Production at HERMES}\label{sec4} 
%
Single-inclusive hadron photoproduction can be also studied with the 
{\sc Hermes} experiment at DESY where an electron (or positron) beam
with $E_e\simeq$~27.5~GeV is scattered off a proton or deuterium
gas target. The available c.m.s.\ energy of about $\sqrt{S}\simeq 7.5$~GeV
is lower than at {\sc Compass} which even further limits the range
of accessible transverse momenta.
On average the electron beam polarization is ${\cal{P}}_{e}\simeq 53\%$.
For the polarization of the gas target we take ${\cal{P}}_d\approx{\cal{P}}_p\simeq 85\%$,
and, contrary to a solid-state target, there is no dilution, 
i.e., ${\cal{F}}_p={\cal{F}}_d=1$.
To estimate the statistical accuracies for a measurement of $A_{\mathrm{LL}}^{\pi,p}$ and
$A_{\mathrm{LL}}^{\pi,d}$ below, we assume an integrated luminosity ${\cal{L}}$ of 
$50\,\mathrm{pb}^{-1}$ and $200\,\mathrm{pb}^{-1}$ for proton and deuterium
targets, respectively. These numbers are based on the actual data already collected
with the {\sc Hermes} experiment which is now running with transverse 
polarization most likely until the end of their experimental program.
%
%
\begin{figure}[t]
\begin{center}
\vspace*{-.5cm} 
\includegraphics[width=0.49\textwidth,clip=]{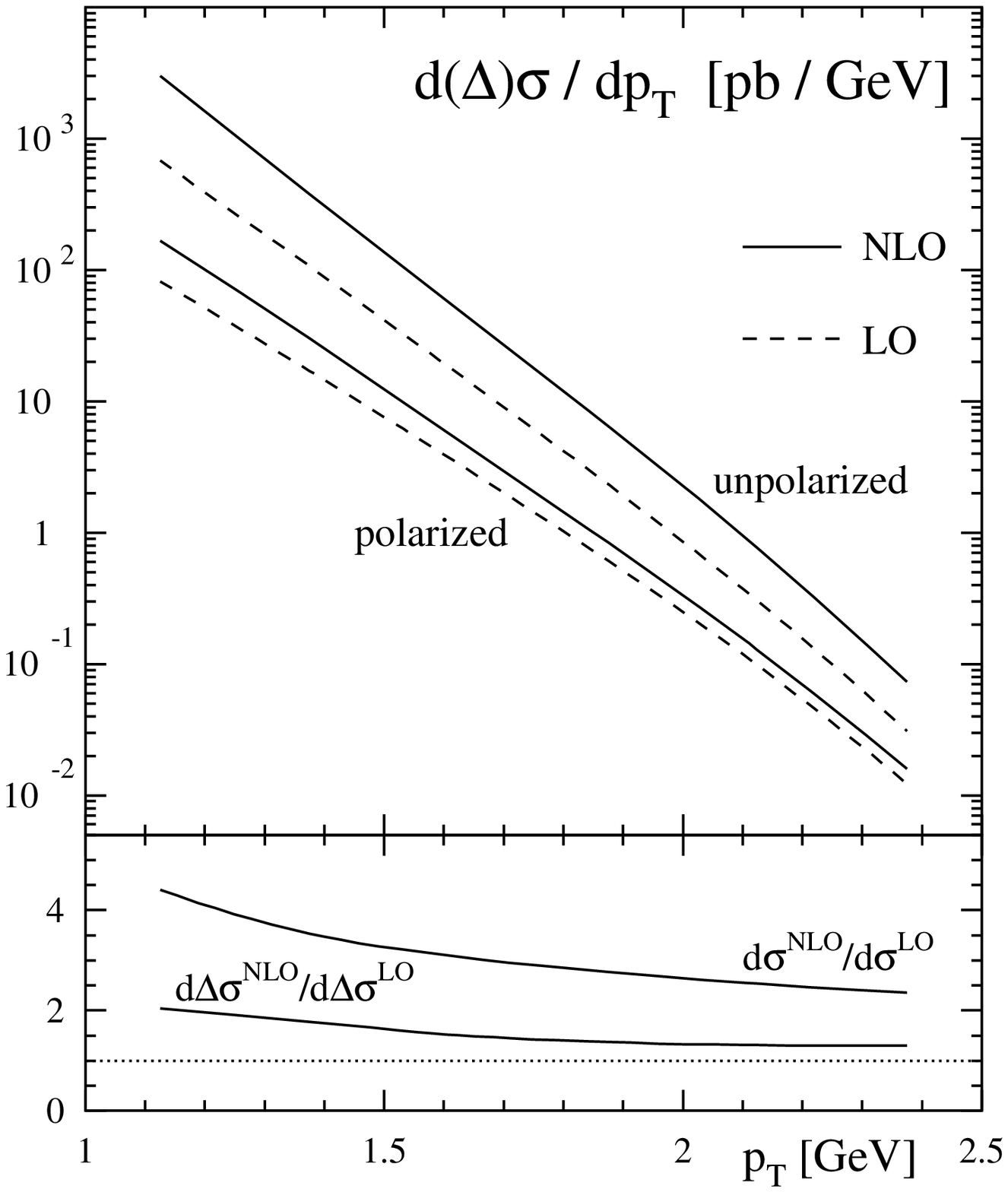}     
        \caption{Unpolarized and polarized $p_T$-differential
                single-inclusive cross sections at LO (dashed) and NLO (solid) for
                photoproduction of neutral pions, $e p\to e'\pi^0 X$, at 
                $\sqrt{S}=7.5$~GeV, integrated over the angular acceptance of
                {\sc Hermes}. 
                The lower panel shows the ratios of NLO to LO contributions
                ($K$-factor).} \label{fig:hermes-xsec-kfac} 
\vspace*{-.35cm} 
\includegraphics[width=0.49\textwidth,clip=]{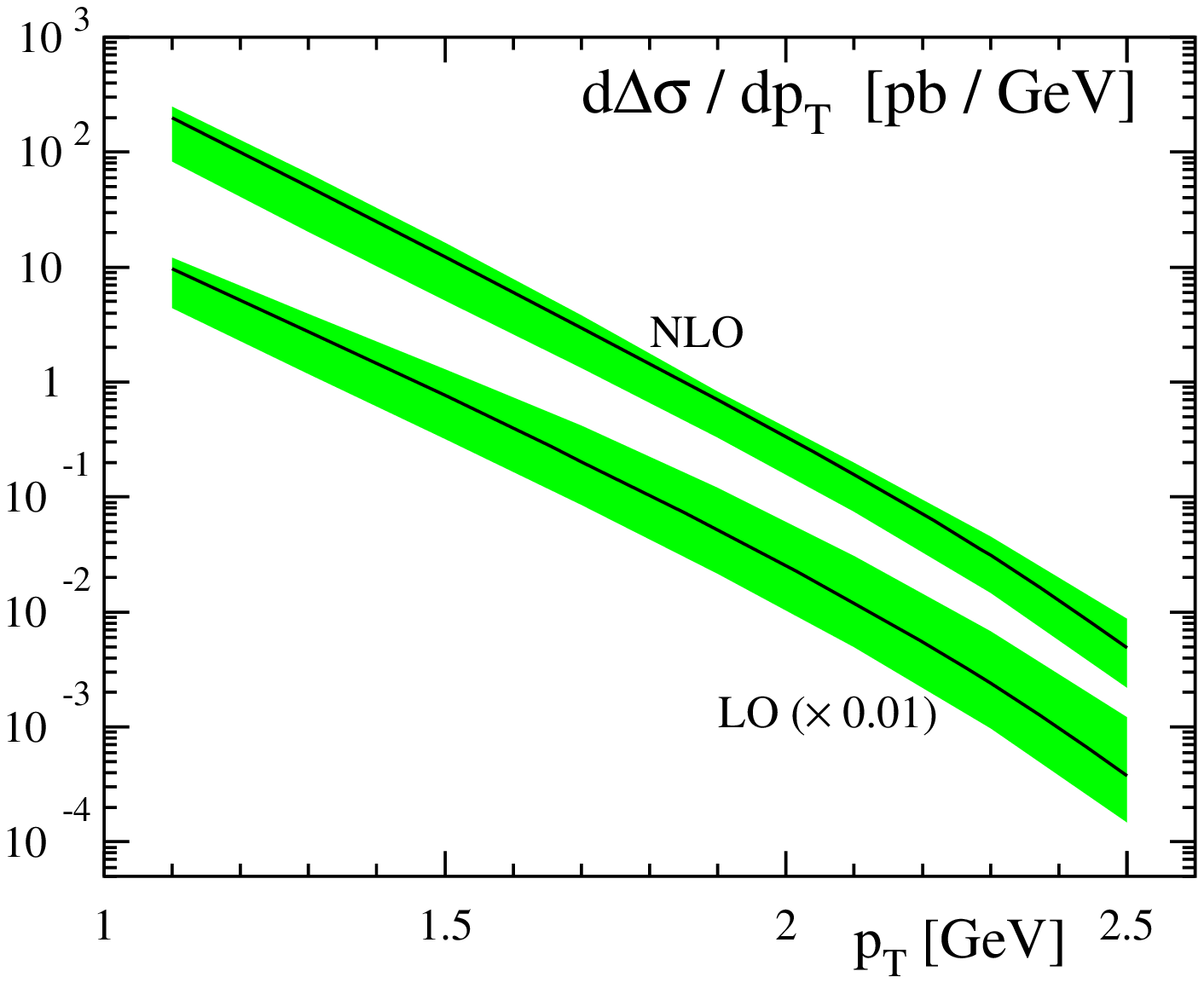}     
        \caption{As in Fig.~\ref{fig:compass-scdep} (b) but now for   
                 $e p\to e' \pi^0 X$ at {\sc Hermes}.
              } \label{fig:hermes-scdep}  

\end{center}
\end{figure}
%
%

%
\begin{figure}[t]
\begin{center}
\vspace*{-.5cm} 
\includegraphics[width=0.49\textwidth,clip=]{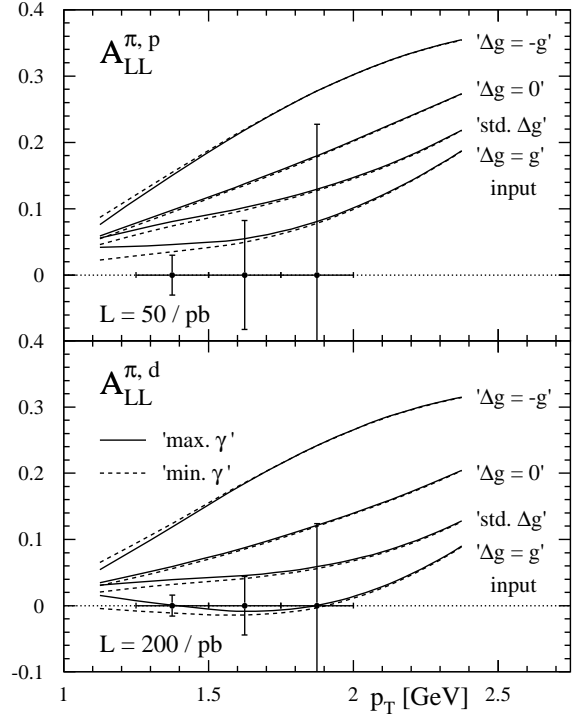}     
        \caption{Double-spin asymmetries $A_{\mathrm{LL}}^{\pi,p}$ and
         $A_{\mathrm{LL}}^{\pi,d}$ for {\sc Hermes} kinematics at NLO for 
         different gluon polarizations in the
         nucleon and minimal and maximal saturation of the polarized
         photon densities, dashed and solid lines, respectively.
         The ``error bars'' indicate the estimated statistical 
         uncertainty assuming an integrated luminosity of
         $50\,\mathrm{pb}^{-1}$ and $200\,\mathrm{pb}^{-1}$
         for $A_{\mathrm{LL}}^{\pi,p}$ and $A_{\mathrm{LL}}^{\pi,d}$, respectively.}
        \label{fig:hermes-asy}     
\end{center}
\end{figure}
%
%
The acceptance of the {\sc Hermes} experiment allows the
detection of pions in the angular range 40~mrad $\lesssim \theta\lesssim 220$~mrad. 
The upper limit on $\theta$ corresponds to a minimal c.m.s.\ rapidity of
$\eta_{cms}^{\min}\simeq 0.2$. At small $p_T$ the lower limit on $\theta$, i.e.,
$\eta_{cms}^{\max}\simeq 1.91$, sets a more stringent bound on $\eta_{cms}^{\max}$
than purely kinematical considerations, $\eta_{cms}^{\max}=\cosh^{-1}({\sqrt{S}/2p_T})$.
For all our numerical studies we integrate over rapidity as before.
We select photoproduction by demanding that the virtuality squared of the photon is less 
than $Q^2_{\max}=0.01\,\mathrm{GeV}^2$. The photon flux with photon energies limited to the
range $0.2\leq y\leq 0.9$ is again modeled by Eq.~(\ref{eq:weiz-will}). 

%
%
\begin{figure}[tp]
\begin{center}
\vspace*{-.5cm} 
\includegraphics[width=0.49\textwidth,clip=]{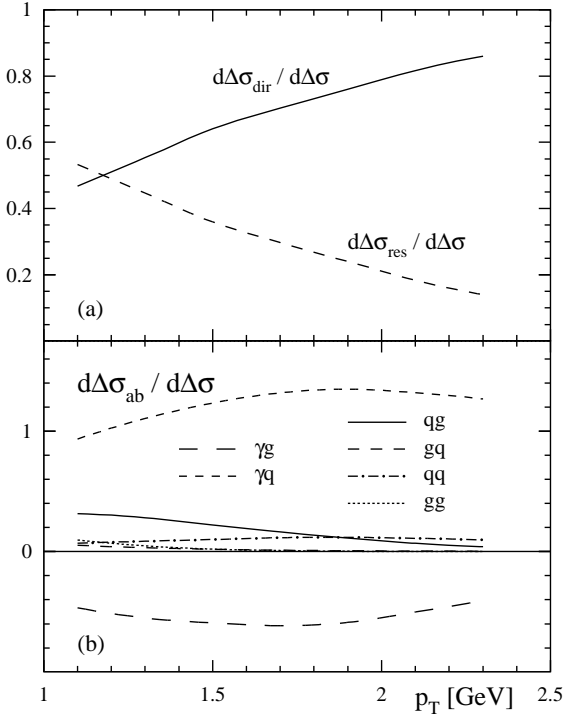}     
        \caption{{\bf{(a)}} Direct and resolved NLO($\overline{\mathrm{MS}}$)
         and {\bf {(b)}} NLO($\overline{\mathrm{MS}}$) contributions
         of different partonic channels $a+b \to c+X$ to the full NLO
         polarized photoproduction cross section shown in Fig.~\ref{fig:hermes-xsec-kfac}.}
         \label{fig:hermes-subproc}
\end{center}
\end{figure}
%
%
%
The resulting polarized and unpolarized $p_T$-differential cross sections and $K$-factors for
photoproduction of neutral pions are depicted in Fig.~\ref{fig:hermes-xsec-kfac}. 
We only show results for a {\sc Hermes} setup with a proton target, i.e.,
$e p\to e' \pi^0 X$, as results
for a deuterium target are qualitatively very similar.
Due to the small c.m.s.~energy available the experimentally accessible transverse momenta 
$p_T$ are limited to a region where a successful and reliable application of perturbative methods
cannot be taken for granted. This is emphasized, for instance, by the large value of the 
unpolarized $K$-factor which ranges from 2 to 4. Even more dramatic are the
uncertainties related to the choice for the renormalization and factorization scales
which are shown for the polarized photoproduction cross section in Fig.~\ref{fig:hermes-scdep}.
Again we have varied all scales simultaneously in the range
$p_T/2\leq\mu_r=\mu_f=\mu_f'\leq 2 p_T$. There is only a slight improvement
when the NLO QCD corrections are taken into account.

These results again clearly indicate that special care has to be taken in the interpretation
of upcoming results in terms of pQCD and, ultimately, of $\Delta g$ at 
current fixed-target experiments.
It is perhaps even more important than at {\sc Compass} energies
that studies of the theoretically better predicted unpolarized cross section
accompany studies with polarization. All reservations notwithstanding,
we analyze in Fig.~\ref{fig:hermes-asy} the sensitivity of hadron 
photoproduction at {\sc Hermes} to the gluon polarization of the nucleon.
We show the double spin asymmetries $A_{\mathrm{LL}}^{\pi,p}$ 
and $A_{\mathrm{LL}}^{\pi,d}$ at NLO for different assumptions about the gluon
polarization of the nucleon and the polarized parton densities of the photon.
Again, we estimate the statistical error for such a measurement with the help
of Eq.~(\ref{eq:all-error}) with the parameters as specified at the
beginning of this section. Clearly, meaningful results about $\Delta g$
can be only obtained for $p_T\approx 1.5\,\mathrm{GeV}$.
Due to the larger data sample available, results obtained with the deuteron target
are much more discriminating than with the proton target.

Figure~\ref{fig:hermes-asy} also shows that the uncertainties from 
estimating the resolved contribution to the cross section are not very
pronounced even at the lowest $p_T$-values shown.
To examine the importance of the resolved contribution at {\sc Hermes}
further, we show in Figs.~\ref{fig:hermes-subproc} (a) and (b) the
relative contributions of the direct and resolved and different
partonic subprocesses $a+b\rightarrow c+X$ to the full photoproduction
cross section, respectively.
It turns out that for a proton target the direct photon contribution
$d\Delta\sigma_{\mathrm{dir}}$ dominates in the entire $p_T$-range.
Contrary to the situation for {\sc Compass}, cf.\ Figs.~\ref{fig-comp-subproc}
(a) and (b), the cancellation of the
two direct channels $\gamma g$ and $\gamma q$ in the sum $d\Delta\sigma_{\mathrm{dir}}$
is less complete.
All resolved subprocesses are small and, for the same reasons
as discussed before for {\sc Compass}, the bulk of $d\Delta\sigma_{\mathrm{res}}$
stems from the scattering of a quark carrying a large momentum fraction of
the photon off a gluon in the nucleon [solid line in Fig.~\ref{fig:hermes-subproc} (b)].
For a deuteron target one obtains qualitatively similar results. However,
at the small $p_T$-values shown the resolved contribution is more
important due to a more pronounced cancellation of the $\gamma g$ and $\gamma q$
subprocesses in $d\Delta\sigma_{\mathrm{dir}}$.

\subsection{Comparison to the E155 Data}\label{sec3e155} 
%
As we mentioned in the introduction, there is already a data
set on the spin asymmetry in single-inclusive hadron production
by the {\sc E155} collaboration~\cite{ref:e155}, and it is of course 
interesting to compare our NLO calculations to these data. The main data 
set of {\sc E155} is for $e N\rightarrow H^{\pm} X$, where $H^+$ ($H^-$)
denotes any hadron of identified positive (negative) charge. Only for the 
data at the larger scattering angle $\theta=5.5^\circ$ are the observed 
hadron transverse momenta large enough for a sensible comparison to
pQCD hard-scattering. Here we use again the various 
GRSV sets of polarized parton distributions, and the ``maximal'' set
of polarized photon densities for the resolved contribution. We use
the fragmentation functions of Ref.~\cite{ref:kretzer} which provide
separate sets of $H^+$ and $H^-$ fragmentation functions. Finally,
because the scattered electron was not observed for the {\sc E155} 
data, we use the Weizs\"{a}cker-Williams spectrum in Eq.~(\ref{eq:weiz-will})
in the case when all photon virtualities are integrated, and we do not
impose a cut on the variable $y$. 

\begin{figure}[tp]
\begin{center}
\vspace*{-.5cm} 
\includegraphics[width=0.49\textwidth,clip=]{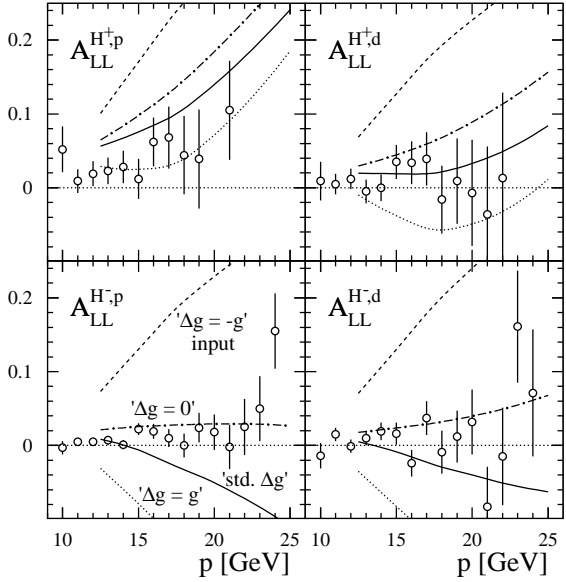}     
        \caption{Comparison of our NLO calculations to the 
{\sc E155} data for the double-spin asymmetries for $H^+$ and $H^-$ 
production in electron scattering off proton and deuteron targets
at scattering angle $\theta=5.5^\circ$ \cite{ref:e155}.
In each of the four panels,
different line styles distinguish the different GRSV sets of polarized
parton distributions used in the calculation.}
         \label{fig:e155}
\end{center}
\end{figure}

Figure~\ref{fig:e155} shows the comparison of our NLO calculations
to the data of {\sc E155} for $H^+$ and $H^-$ production off proton
and deuteron targets. One can see that it is difficult to achieve 
a quantitative overall agreement for a given set of GRSV polarized
parton densities. On the other hand, the results show very large 
variation with the polarized parton distributions, and fine details
of the densities will therefore matter. This is in particular
so as there are again strong cancellations between the direct and 
resolved contributions, and even within these. One should also keep in mind
that the uncertainty in the charge separation of the available
sets of $H^+$ and $H^-$ fragmentation functions is currently 
very hard to quantify and certainly large. We finally note that there is
also room for the ``soft'' non-perturbative contributions 
to the asymmetry discussed in~\cite{ref:afanasev1}, even though
the data do not give compelling evidence for their presence
either.
%
%
%

\section{Discussion and Conclusions}\label{sec5}
%
To summarize, in this paper we have presented a detailed 
phenomenological study of single-inclusive
hadron photoproduction at fixed-target energies with particular
emphasis on the kinematics relevant for the {\sc Compass} and {\sc Hermes}
experiments. We have also compared our calculations to already
available data from the {\sc E155} experiment.
All calculations were consistently performed at the next-to-leading
order of perturbative QCD.
We have carefully explored and critically discussed the
applicability of the perturbative methods in the region of relatively
low momentum transfers characterized by the $p_T$-values accessible
in fixed-target experiments. Both the $K$-factors as well as uncertainties
due to the choice of the renormalization and factorization scales turn
out to be sizable; the latter only slightly improving when NLO corrections
are taken into account. 

All this indicates that the pQCD framework in the {\sc Compass}
and {\sc Hermes} kinematic regimes is delicate, and calls for
a ``reference measurement'' of the unpolarized photoproduction cross section
before the spin asymmetry can be analyzed in terms of $\Delta g$.
If the unpolarized cross sections turn out to be in reasonable agreement
with theory, measurements of the corresponding spin asymmetries 
by {\sc Compass} and {\sc Hermes} will add valuable information
about the gluon polarization of the nucleon at momentum fractions
$x \simeq 0.1 \div 0.3$. To access these large $x$-values at
a polarized $pp$ collider like RHIC requires to detect final-state hadrons, photons,
or jets at rather high $p_T$ where the rate is small. The theoretical 
analysis of the unpolarized cross sections at {\sc Compass} and {\sc Hermes} 
may well require to pursue avenues like performing all-order resummations of
large logarithms in the perturbative series. 

Finally, it turned out that the resolved photon contribution to the
cross section is sizable, and, under certain conditions, even dominant, 
but independent of the details of the non-perturbative spin structure 
of the polarized photon which is completely unknown so far. 
Analyzing the data with the approximation $d\Delta\sigma_{\mathrm{res}}\approx 0$
would lead to incorrect conclusions about $\Delta g$.
To access the parton content of polarized photons certainly requires
a future polarized lepton-hadron collider facility such as the
\mbox{eRHIC} project at BNL.

\section*{Acknowledgments}
%
We are grateful to E.\ Aschenauer, P.\ Liebing, and V.\ Mexner ({\sc Hermes}), to
J.\ Friedrich, R.\ Kuhn, J.-M.\ Le Goff, G.\ Mallot, S.\ Paul, and 
J.\ Pretz ({\sc Compass}), and to P.\ Bosted ({\sc E155}) for valuable 
discussions and information about the experimental details of their photoproduction 
measurements. We also thank A.\ Afanasev for drawing our attention to
the {\sc E155} data and to his earlier work.
W.V.\ is grateful to RIKEN, Brookhaven National Laboratory and the U.S.\
Department of Energy (contract number DE-AC02-98CH10886) for
providing the facilities essential for the completion of this work. B.J.\ would
like to thank the Sonderforschungsbereich TR9 of the Deutsche
Forschungsgemeinschaft for support. 


\end{document}